\documentclass{article}
\usepackage{geometry}
\usepackage{authblk}
\usepackage{amsmath}
\usepackage{amssymb}
\usepackage{graphicx}

\begin{document}
\title{Quantum continual learning of quantum data realizing knowledge backward transfer}
\author[1]{Haozhen Situ}
\author[1]{Tianxiang Lu}
\author[3,4]{Minghua Pan}
\author[,2]{Lvzhou Li\thanks{Corresponding author.\\ E-mail addresses: lilvzh@mail.sysu.edu.cn}}
\affil[1]{\scriptsize College of Mathematics and Informatics, South China Agricultural University, Guangzhou 510642, China}
\affil[2]{Institute of Quantum Computing and Computer Theory, School of Computer Science and Engineering, Sun Yat-sen University, Guangzhou 510006, China}
\affil[3]{Guangxi Key Laboratory of Cryptography and Information Security, Guilin University of Electronic Technology, Guilin 541004, China}
\affil[4]{Department of Physics, Tsinghua University, Beijing 100084, China}

\date{}
\maketitle

\begin{abstract}
For the goal of strong artificial intelligence that can mimic human-level intelligence, AI systems would have the ability to adapt to ever-changing scenarios and learn new knowledge continuously without forgetting previously acquired knowledge. When a machine learning model is consecutively trained on multiple tasks that come in sequence, its performance on previously learned tasks may drop dramatically during the learning process of the newly seen task. To avoid this phenomenon termed catastrophic forgetting, continual learning, also known as lifelong learning, has been proposed and become one of the most up-to-date research areas of machine learning. As quantum machine learning blossoms in recent years, it is interesting to develop quantum continual learning. This paper focuses on the case of quantum models  for quantum data  where the computation model and  the data to be processed  are both quantum.  The gradient episodic memory method is incorporated to design a quantum continual learning scheme that overcomes catastrophic forgetting and realizes knowledge backward transfer. Specifically, a sequence of quantum state classification tasks is continually learned by a variational quantum classifier whose parameters are optimized by a classical gradient-based optimizer.
The gradient of the current task is projected to the closest gradient, avoiding the increase of the loss at previous tasks, but allowing the decrease. Numerical simulation results show that our scheme not only overcomes catastrophic forgetting, but also realize knowledge backward transfer, which means the classifier's performance on previous tasks is enhanced rather than compromised while learning a new task.
\end{abstract}

\section{Introduction}

Promoted by the interaction between quantum mechanics and computer science, quantum computation \cite{QCQI} is a brand new computation scheme that manipulates quantum information units following the rules of quantum mechanics. Due to the exponentially increasing Hilbert space and unique properties like superposition and entanglement, quantum computation possesses tremendous potential computing power. Many prominent research achievements have appeared in the realm of quantum computation since the end of last century. In recent years, quantum computation has received high attention and significant investment from the industry.

One of the research focuses in quantum computation is quantum machine learning (QML) \cite{QML}, which is the mergence between quantum computation and machine learning \cite{ML}. The efficiency of machine learning can be enhanced by quantum algorithms such as the quantum algorithm for linear systems of equations \cite{HHL}, quantum algorithm for data fitting \cite{datafitting}, quantum support vector machine \cite{QSVM,QTSVM}, quantum principal component analysis \cite{QPCA} and quantum recommendation system \cite{QRS}. On the other hand, using quantum circuits with learnable parameters (a.k.a. variational quantum circuits or parameterized quantum circuits) as ML models is a feasible way to integrate quantum and classical computation in the NISQ \cite{NISQ} era. Two typical pioneering quantum-classical hybrid algorithms are variational quantum eigensolver \cite{VQE} and quantum approximate optimization algorithm \cite{QAOA}. Various common ML problems have been considered using variational quantum algorithms such as classification \cite{classifyHCT,classifySK,QCNN,reupload,classifyPCH}, generalization \cite{genBGP,genSitu,genZLW,genCHL,genRA,genHOR} and autoencoder \cite{aeROA,aeLAM,aePTP,aeHMY}. The structure of the variational quantum circuit (layout of gates) is fixed and the parameters of the gates are optimized at the first place. Recently, variational quantum compiling \cite{vqcKLP,vqcZZZ,vqcHLZ} and quantum architecture search \cite{qasZHZ,qasLSD} try to optimize the structures of variational quantum circuits to obtain more efficient circuits.

Intelligence creatures have the ability to continually learn new knowledge and skills and incrementally adapt to new scenarios over the duration of their lifetime. However, learning of new tasks for an ML model often results in forgetting of old tasks, which is called the catastrophic forgetting phenomenon. Continual learning (a.k.a lifelong learning, incremental learning, sequential learning) \cite{CL} is a research topic that provides solutions for acquiring knowledge from  continuously changing data stream of multiple tasks. Recently, continual learning has been explored in the context of quantum machine learning \cite{QEWC}. It's been shown that a quantum classifier's performance on old tasks may deteriorate when learning a new task. On the assumption that some parameters in variational quantum classifiers are more important than others, the elastic weight consolidation (EWC) \cite{EWC} method can help to protect those parameters with great importance to the old tasks from being updated drastically. Numerical experiments demonstrate that a quantum classifier can continually learn three different classification tasks without catastrophic forgetting. Another goal of continual learning is to learn a new task faster by utilizing past knowledge acquired from old tasks. Ref. \cite{CRL} proposed a continual reinforcement learning framework to deal with the constantly changing environment. In this framework, the agent leverages past policies learned from previous noise environments to generate a state preparation quantum circuit for a new noise environment. Without training from scratch, the convergence speed can be much faster.

In this work, we investigate the continual learning of a sequence of quantum state classification tasks. The training and testing data are all quantum states, challenging to classical neural networks because the  exponential increasing resources required to represent them classically. We adopt the quantum-classical hybrid computing scheme to train a variational quantum classifier, the parameters of which are updated by a classical gradient descent optimizer. We incorporate the gradient episodic memory (GEM) method \cite{GEM} to keep the classifier remembering previously learned tasks. Before the gradient is used to update the parameters, it's modified to avoid the reduction of the classifier's performance on old tasks. A merit of this method is the possibility of positive backward transfer, which denotes the benefits previous tasks get from the training of new tasks. Numerical results demonstrate that the continual learning of six quantum state classification tasks with GEM has higher accuracy and backward transfer than EWC.

This paper is organized as follows. In section \ref{sec:related}, we present the related work of classical and quantum continual learning. In section \ref{sec:method}, we describe the quantum state classification tasks, the structure of our variational quantum classifier and the GEM method. In section \ref{sec:results}, the numerical experiment results of quantum continual learning with GEM and other strategies are shown and compared. Section \ref{sec:conclusion} presents the concluding remarks.

\section{Related work}
\label{sec:related}

Elastic weight consolidation (EWC) \cite{EWC} is a famous continual learning method overcoming catastrophic forgetting phenomenon in neural networks. It adds a penalty term to the loss function to confine the parameter update, because the deviation of important parameters from the optimal solutions of previously trained tasks will compromise the classifier's performance on those tasks. More specifically, the diagonal elements of the Fisher information matrix at the optimal point are saved when each task has been trained. This information tells about the importance of the parameters to the task and can be used to constrain the update strength of parameters. When training the $t$-th task, the model's parameter $\theta$ is optimized to minimize the following loss function
\begin{align}
L(\theta) = L^t(\theta) + \lambda \sum_{k<t}\sum_j F^k_j (\theta_j - \theta^k_j)^2,
\end{align}
where $L^t(\theta)$ is the original loss function for the $t$-th task, $\lambda$ is a hyper-parameter  determining how important the old tasks are compared with the new one, $F^k_j$ is the $j$-th diagonal element of the Fisher information matrix at the optimal point $\theta^k$ for the $k$-th task. By adding the quadratic penalty term, each parameter is pulled back toward its old values by an amount proportional to its importance for performance on previously learned tasks.

The gradient episodic memory (GEM) method \cite{GEM} uses samples of previous tasks kept in the memory to modify the gradient of the current task. Unlike EWC, the performance of the old tasks is more likely to increase because the previous samples are rehearsed in the parameter update. This phenomenon is called positive backward transfer. Ref. \cite{GEM} proposed an evaluation metric
\begin{align}
\mathrm{BWT} = \frac{1}{N_t-1} \sum_{i=1}^{N_t-1} R_{N_t, i} - R_{i,i}
\end{align}
to measure the influence that learning a task has on the performance on a previous task. $N_t$ denotes the number of tasks. $R_{i,j}$ denotes the test accuracy of the model on the $j$-th task after learning the $i$-th task.
Positive BWT means learning new tasks increases the performance of old tasks, and negative BWT means learning new tasks decreases the performance of old tasks, i.e., catastrophic forgetting. Another evaluation metric we use in this paper is the average accuracy defined by
\begin{align}
\mathrm{ACC} = \frac{1}{N_t} \sum_{i=1}^{N_t} R_{N_t, i},
\end{align}
which is the average accuracy on all tasks after the last task has been learned. This is the most commonly used metric in continual learning.

Continual learning has been investigated in the QML field. Ref. \cite{QEWC} adopted the EWC method to help a quantum classifier overcome catastrophic forgetting and learn a sequence of three classification tasks successfully. Ref. \cite{CRL} studied quantum state preparation for changing environment by training a reinforcement learning agent that utilizes past policies learned from previous environments.

Another subject called transfer learning, which is similar to continual learning to some extent, has also been studied in the QML field. In transfer learning, a model is trained on task B taking advantage of the knowledge of a similar task A. Unlike continual learning, the model's performance on only task B is concerned. A generic quantum transfer learning scheme was considered in Ref. \cite{TL-Mari}. In this scheme, a pre-trained classical or quantum model for task A is truncated by removing some final layers. A new trainable classical or quantum model is connected to the end of the pre-trained model. While keeping the parameters of the pre-trained model constant, the new model's parameters are optimized using the dataset of task B. Ref. \cite{TL-Wang} proposed a quantum deep transfer learning scheme for feature-based knowledge transfer, which is composed of a model for generating quantum feature data based on quantum Boltzmann machine \cite{QBM}, an algorithm for computing kernel matrices of mixed quantum states, and a quantum alignment algorithm for comparing the characters of the kernel matrices.

\section{Methods}
\label{sec:method}

We use a variational quantum classifier to classify quantum states in the continual learning fashion. As far as we know, it's the first time a quantum classifier is trained to solve up to six quantum state classification problems. Each of the six tasks contains a dataset of quantum states that belong to the positive class or the negative class. After random initialization of parameters, the classifier reads the training data from one task after another. The classifier switches to the next task only when it has read all training samples of the current task.  After being trained through all the tasks, the classifier is expected to have good performance on all tasks, i.e., predict the class label for a testing quantum state corresponding to anyone of the trained tasks as accurately as possible.

\subsection{Quantum state classification tasks}
\label{sec:tasks}

We use six quantum state classification tasks as benchmark. Task 1 is to classify the symmetry protected topological phases for the following Hamiltonian:
\begin{align}
H(h) = - \sum_{i=1}^{N_q} \sigma_x^{i-1} \sigma_z^{i} \sigma_x^{i+1} + h \sum_{i=1}^{N_q} \sigma_y^i \sigma_y^{i+1},
\end{align}
where $N_q$ is the number of spins, $\sigma_{x,y,z}^i$ are the Pauli operators acting on the $i$-th spin and $h$ is the strength of the nearest-neighbor interaction. This model features a quantum phase transition at $h=1$, between a $\mathbb{Z}_2\times\mathbb{Z}_2$ symmetry protected topological phase characterized by a string order for $h<1$, and an antiferromagnetic phase with long-range order for $h>1$. The ground states of $H$ comprise the dataset of task 1. We set $N_q=8$ and obtain 512 samples by picking $h$ evenly from 0 to 2.

Ref. \cite{NTangled} provided an entangled dataset for quantum machine learning, which contains quantum states with different amounts of multipartite entanglement. Five classes of 8-qubit states with concentratable entanglement (CE) 0.1, 0.15, 0.25, 0.4 and 0.45 are given. Here we choose the classification between states with CE=0.1 and states with CE=0.25 as task 2, and the classification between quantum states with CE=0.15 and 0.45 as task 3. Each dataset of these two tasks contains 512 quantum states.

The remaining three tasks consider quantum states evolved under the time-dependent two-body transverse field Ising model:
\begin{align}
H(\tau,J) = (1-\tau) \sum_{i=1}^{N_q} \sigma_x^i + \tau \sum_{i<j}J\sigma_z^i\sigma_z^j.
\end{align}
We let $N_q=8$ and make 512 quantum states $e^{-iH}|+\rangle^{\otimes N_q}$ by picking $J$ evenly from -1 to 1,  with fixed $\tau\in\{0.25, 0.5, 0.75\}$ for task 4, 5, 6, respectively. The class label is dependent on the sign of $J$.

The order of the above tasks in which they are fed to the classifier can be denoted by a sequence, e.g. 123456. In section \ref{sec:results} we report the numerical experimental results of different orders. We summarize the six quantum state classification tasks in Table \ref{tab:tasks} for clearness.

\begin{table}
\caption{Six quantum state classification tasks}
\label{tab:tasks}
\centering
\begin{tabular}{cc}
\hline
Task name & Description\\
\hline
Task 1 & Classify the symmetry protected topological phases \\
Task 2 & Classify between quantum states with CE=0.1 and CE=0.25 \\
Task 3 & Classify between quantum states with CE=0.15 and CE=0.45 \\
Task 4 & Classify states evolved under the transverse field Ising model with $\tau=0.25$ \\
Task 5 & Classify states evolved under the transverse field Ising model with $\tau=0.5$ \\
Task 6 & Classify states evolved under the transverse field Ising model with $\tau=0.75$ \\
\hline
\end{tabular}
\end{table}

\subsection{Variational quantum classifier}

The variational quantum classifier in this study adopts the same hardware-efficient variational ansatz as that in Ref. \cite{QEWC}. The quantum circuit ansatz has $N_l$ repeated layers composed of $\mathrm{R_x}$, $\mathrm{R_z}$ and $\mathrm{CNOT}$ gates. Each layer $U_j$ is defined as
\begin{align}
U_j = \bigotimes_{i=1}^{N_q} \mathrm{R_z}^{i}(\alpha_{j,i,3})
\prod_{k=1}^{2}\Bigg(
\bigotimes_{i=1}^{\lfloor\frac{N_q}{2}\rfloor} \mathrm{CNOT}^{2i-1}_{2i}
\bigotimes_{i=1}^{\lfloor\frac{N_q-1}{2}\rfloor} \mathrm{CNOT}^{2i}_{2i+1}\Bigg)
\bigotimes_{i=1}^{N_q} \mathrm{R_z}^{i}(\alpha_{j,i,2})
\bigotimes_{i=1}^{N_q} \mathrm{R_x}^{i}(\alpha_{j,i,1}) ,
\end{align}
where $N_q$ is the number of qubits, $\mathrm{R_{x,z}}^i$ are the rotation gates acting on the $i$-th qubit, the superscript and subscript of $\mathrm{CNOT}$ denote the control and target qubit, respectively.
Before the final measurement, an extra $\mathrm{R_x}$ gate is performed on each qubit, so the effect of the whole circuit can be written as
\begin{align}
|x'\rangle = \bigotimes_{i=1}^{N_q} \mathrm{R_x}^{i}(\beta_i) \prod_{i=1}^{N_l} U_i |x\rangle,
\end{align}
where $|x\rangle$ and $|x'\rangle$ denote the input and output state of the circuit, respectively. The rotation angles $\{\alpha, \beta\}$ are learnable parameters which will be updated by a classical optimizer during the training process.

The quantum classifier in Ref. \cite{QEWC} uses the measurement expectation value of only one specified qubit as the prediction value. In order to utilize more information in the final quantum state, we collect the measurement expectation values of all qubits in the form of a vector $z=(z_1,\ldots,z_{N_q})$, with
\begin{align}
z_i = \langle x'| \sigma_z^i |x'\rangle,
\end{align}
where $\sigma_z^i$ is the observable being measured on the $i$-th qubit.
Then the prediction value is obtained by a postprocessing transform on $z$ through a simple feedforward neural network with only one hidden layer containing $N_q$ neurons. The set of all trainable parameters in our classifier is denoted as $\theta=\{\alpha,\beta,w\}$ with $w$ being the weights of the neural network. The prediction for input $x$ is denoted as $f_\theta(x)$.

\subsection{Gradient episodic memory method}
We incorporate the gradient episodic memory (GEM) method \cite{GEM} for continual learning of quantum state classification tasks. In the training process of the $k$-th task, a fraction of the training samples and labels are allocated to an episodic memory $M_k$. The loss at the episodic memory from the $k$-th task is defined as
\begin{align}
L(f_\theta, M_k) = \frac{1}{|M_k|} \sum_{(x,y)\in M_k} L(f_\theta(x),y).
\end{align}
In this work, $L(f_\theta(x),y)$ is the cross entropy loss which is commonly used for classification problems.

We require that the parameter update in the training process of the $t$-th task would not decrease the classifier's performance on previous tasks. This requirement can be expressed by the following constrained optimization problem:
\begin{align}
& \mathrm{min}_{\theta}\quad L(f_{\theta}(x),y)\nonumber\\
& \mathrm{s.t.}\quad L(f_\theta, M_k) \leq L(f^{t-1}, M_k),\quad \mathrm{for\ all\ } k<t,
\end{align}
where $f^{t-1}$ is the classifier state at the end of learning the $(t-1)$-th task. Let $g=\nabla_\theta L(f_\theta(x),y)$ be the loss gradient of the $t$-th task, and $g_k=\nabla_\theta L(f_\theta, M_k)$ be the loss gradient of $M_k$. If the inner product $\langle g, g_k\rangle \geq 0$ for all $k<t$, then the parameter update according to $g$ is unlikely to increase the loss at previous tasks. Otherwise, the gradient $g$ needs to be projected to the closest gradient $\tilde{g}$, by solving
\begin{align}
& \mathrm{min}_{\tilde{g}}\quad \frac{1}{2}\|g-\tilde{g}\|^2_2\nonumber\\
& \mathrm{s.t.}\quad \langle \tilde{g}, g_k\rangle \geq 0,\quad \mathrm{for\ all\ } k<t,
\end{align}
which is a quadratic program in $p$ variables (the number of parameters in the classifier). In order to solve this efficiently, GEM works in the dual space which results in a much smaller quadratic program with only $t-1$ variables:
\begin{align}
& \mathrm{min}_{v}\quad \frac{1}{2} v^{\top}G^{\top}Gv + g^{\top}Gv  \nonumber\\
& \mathrm{s.t.}\quad v\geq 0,
\end{align}
where $G=-(g_1,\ldots,g_{t-1})\in \mathbb{R}^{p\times (t-1)}$. After the solution $v^{\star}$ is found, the projected gradient can be obtained by $\tilde{g}=Gv^{\star}+g$.

\section{Results}
\label{sec:results}

We conduct numerical simulation experiments on continual learning of six quantum state classification tasks given in section \ref{sec:tasks}. We compare the GEM method and the EWC method, along with plainly training the tasks one by one (denoted as `Plain' in the following). The average accuracy (ACC) and backward transfer (BWT) defined in section \ref{sec:related} are calculated to evaluate the performance.

The simulation is carried out using the Pennylane library \cite{Pennylane} alongside the PyTorch \cite{PyTorch} interface. The stochastic gradient optimizer Adam (Adaptive Moment Estimation) \cite{Adam} is used to update the classifier's parameters with the initial learning rate 0.1. Each task contains 512 samples, 400 of which are used as training samples and the remains are used for testing. The variational quantum classifier has $N_l=1$ layer, and is trained on each task for only 1 epoch with a batch size 10. For the EWC method, 50 samples randomly chosen from the training set are used to calculate the diagonal of the Fisher information matrix. For the GEM method, the capacity of each episodic memory $M_k$ is 50 samples which are randomly picked from the training set of the $k$-th task.

With regard to the training order of the tasks, we first try the lexical order 123456 in which the tasks are introduced in section \ref{sec:tasks}, then we pick another four random orders that can be found in the first column of table \ref{tab:ACCBWT}. For each training order and strategy, we train 5 classifiers with different random initialization, and report the one with the highest test ACC. The results are summarized in table \ref{tab:ACCBWT}. It can be seen that for all task sequences, GEM achieves higher ACC than EWC and plain training. GEM results in comparable ACC higher than $90\%$ for all task sequences except 123456, yet the ACC of EWC is lower than $80\%$ for all task sequences. The BWT is all positive in the case of GEM, but all negative for EWC, and even lower in the case of plain training. We can conclude that the overall performance of GEM is significantly better than EWC and plain training.

\begin{table}
\caption{Average accuracy (ACC) and backward transfer (BWT) for different strategies and task sequences}
\label{tab:ACCBWT}
\centering
\begin{tabular}{ccccccc}
\hline
& \multicolumn{2}{c}{Plain}  & \multicolumn{2}{c}{EWC} & \multicolumn{2}{c}{GEM}\\
Task sequence & ACC & BWT & ACC & BWT &ACC & BWT \\
\hline
123456 & 0.7277 & 0.0268 & 0.7812 & -0.0268 & 0.8482 & 0.0372 \\
234561 & 0.6637 & -0.1458 & 0.7307 & -0.0640 & 0.9524 & 0.1503 \\
436251 & 0.6920 & -0.1443 & 0.7693 & -0.0565 & 0.9390 & 0.0699 \\
312456 & 0.7188 & -0.0714 & 0.7798 & -0.0164 & 0.9554 & 0.1071 \\
246351 & 0.6310 & -0.1815 & 0.7188 & -0.0223 & 0.9405 & 0.1905 \\
\hline
\end{tabular}
\end{table}

\begin{figure}
  \centering
  \includegraphics[width = 12cm]{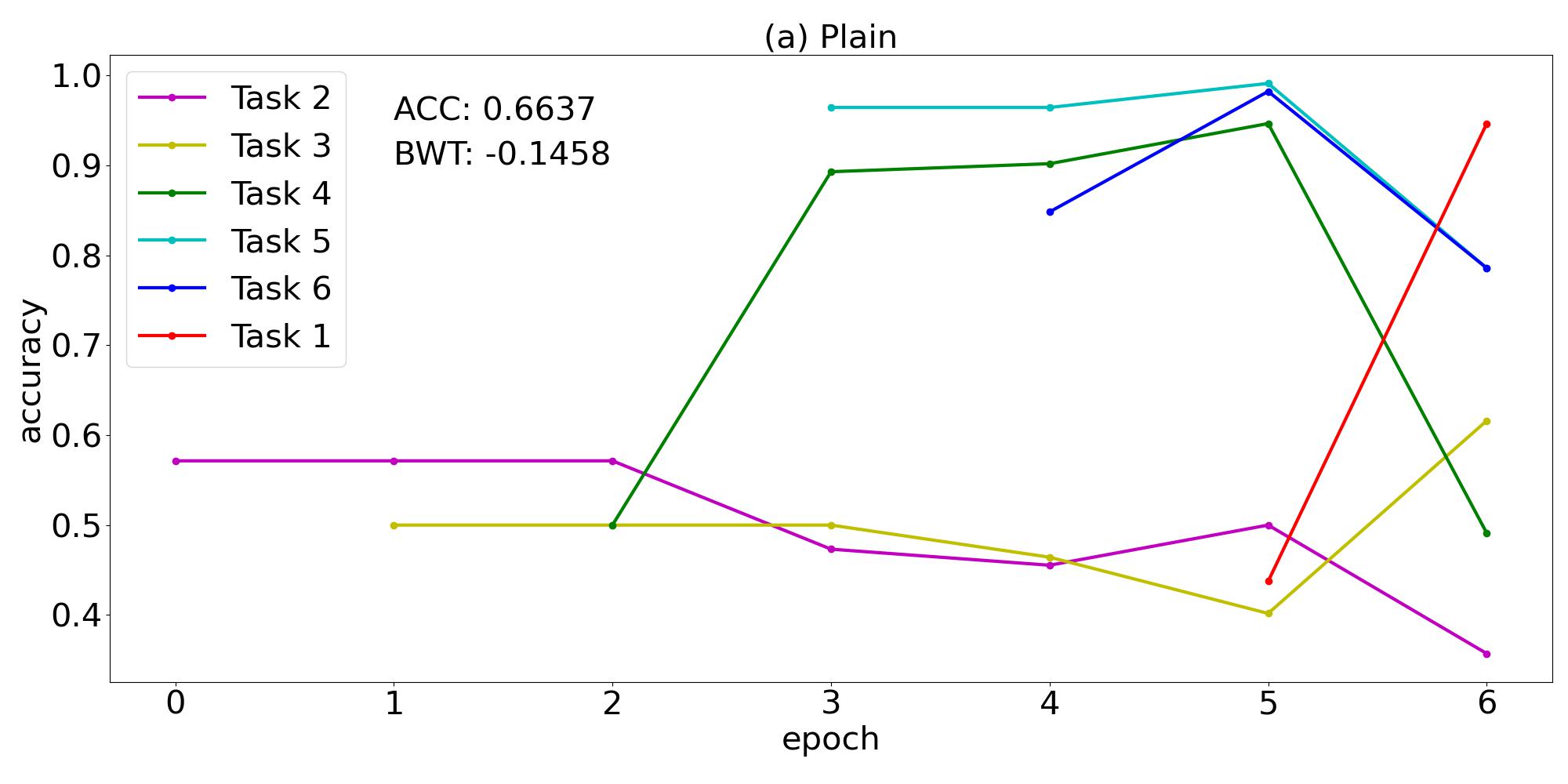}
  \includegraphics[width = 12cm]{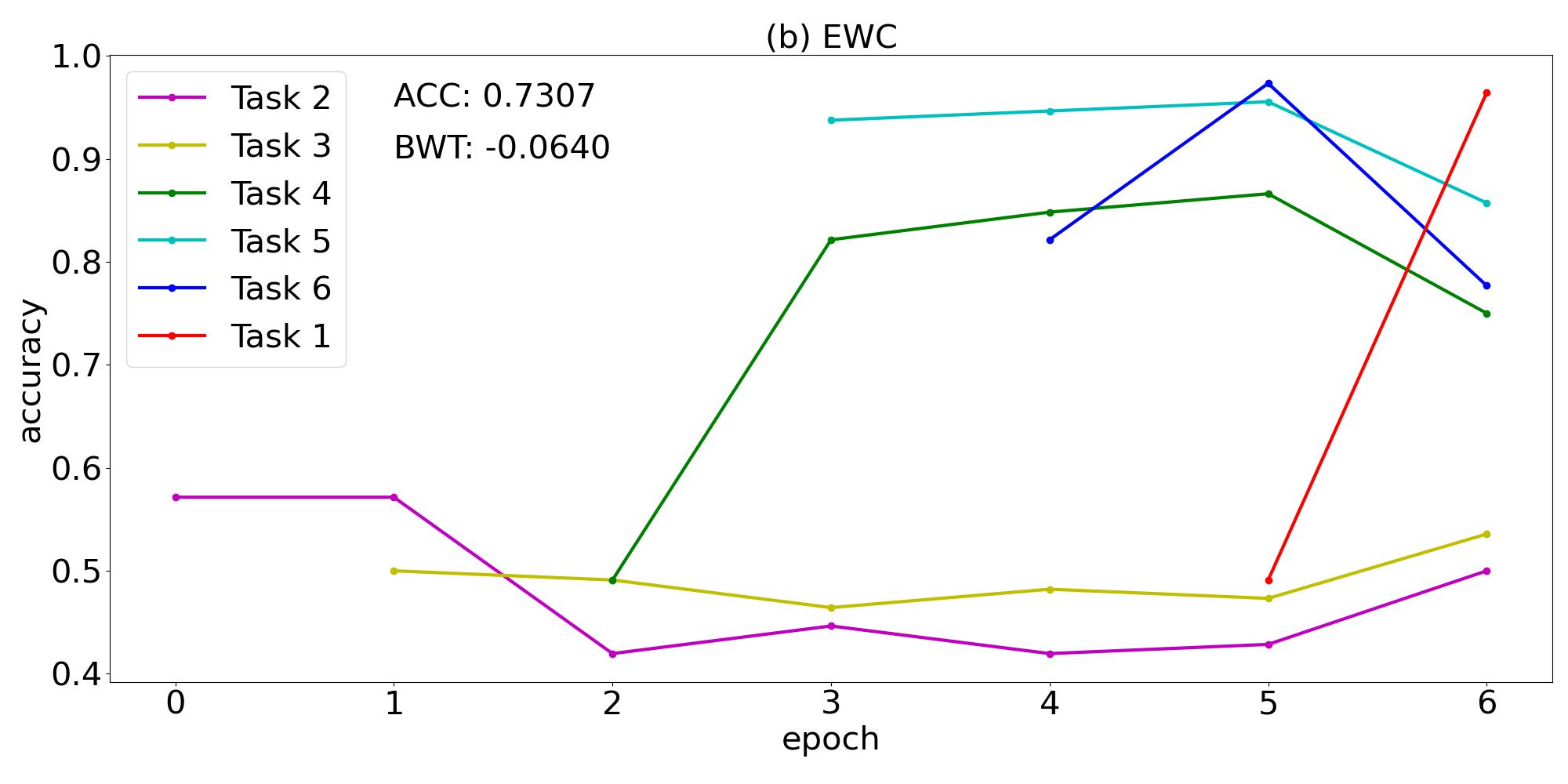}
  \includegraphics[width = 12cm]{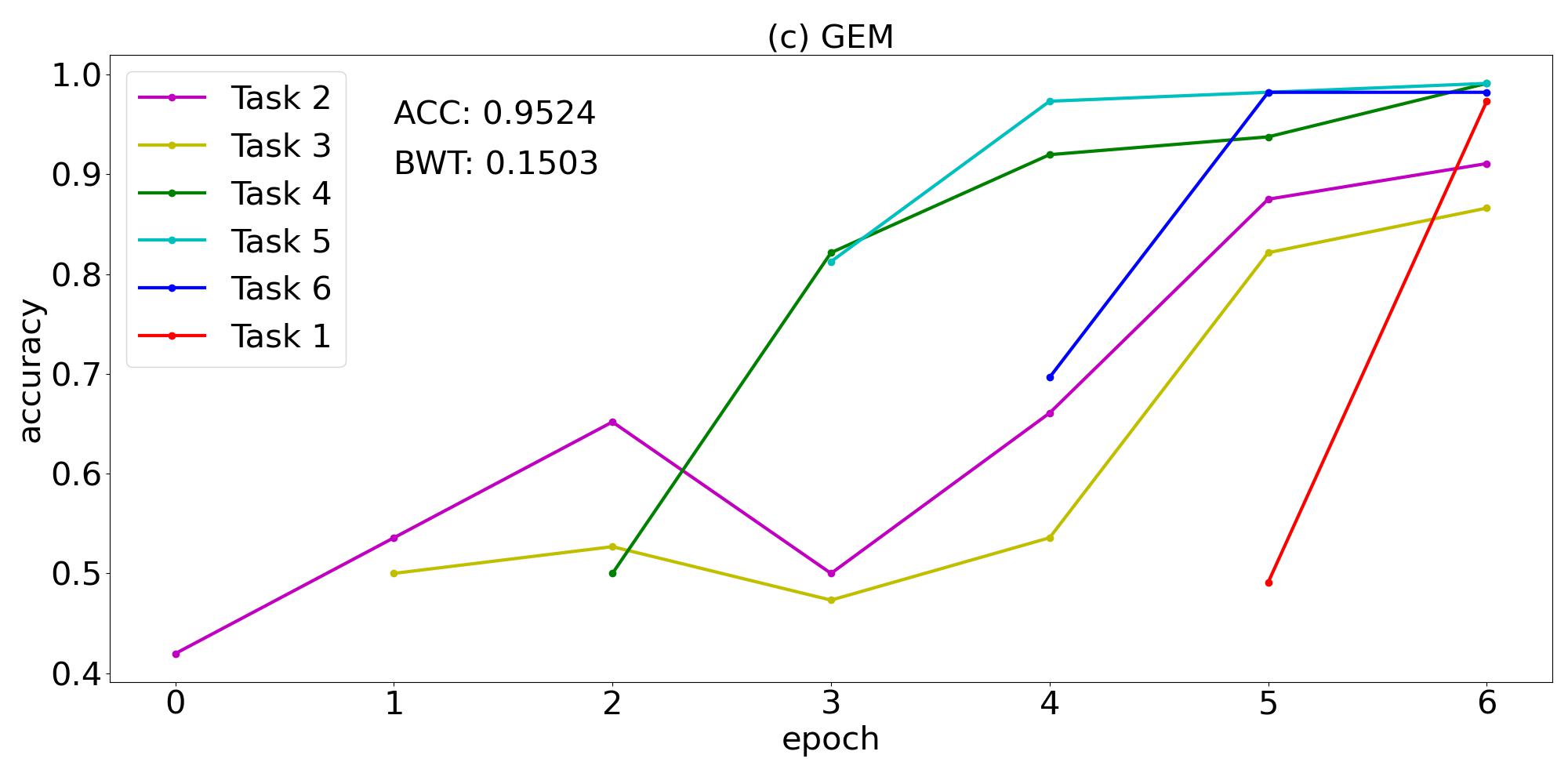}
  \caption{The learning curves for (a) plain training without any continual learning strategy (b) EWC method (c) GEM method.   The task sequence is 234561 as shown in the legends, corresponding to the second row of Table \ref{tab:ACCBWT}. There are 6 epoches in each plot. In the $i$-th epoch, the classifier is trained on the $i$-th task. The testing data of a task is used to evaluate the test accuracy from the moment the training of that task begins until the training of the last task finishes.}
  \label{fig:learningcurve}
\end{figure}

\begin{figure}
  \centering
  \includegraphics[width = 18cm]{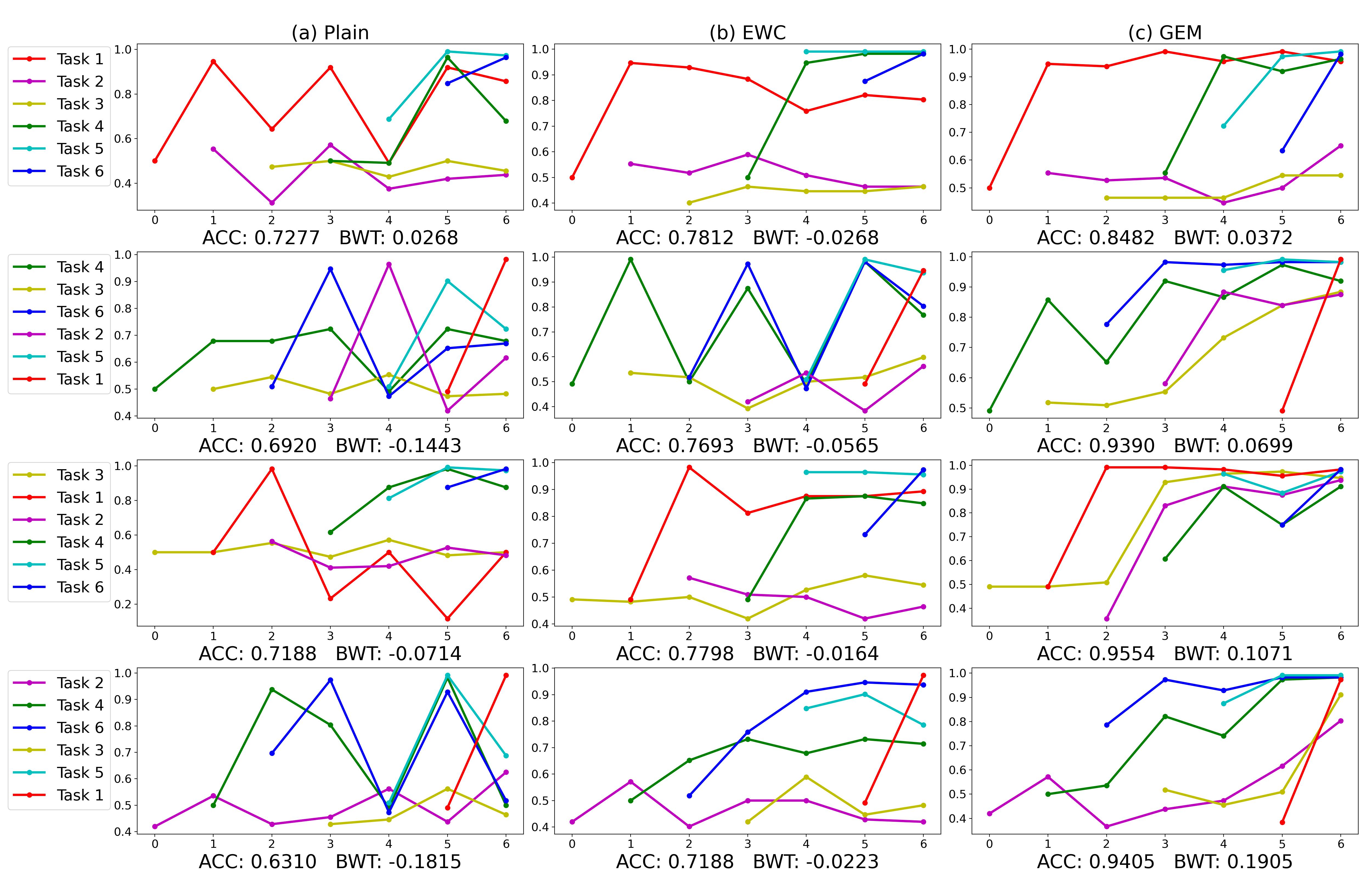}
  \caption{The learning curves for (a) plain training without any continual learning strategy (b) EWC method (c) GEM method. The three plots of each row correspond to the same task sequence shown on the far left. The horizontal and vertical axes represent the epoch and test accuracy respectively.}
  \label{fig:allcurves}
\end{figure}

The training curve of task sequence 234561 is plotted in figure \ref{fig:learningcurve}. The catastrophic forgetting phenomenon appears in figure \ref{fig:learningcurve}(a) which shows the plain training of a sequence of tasks without any continual learning strategy. After the classifier has been trained on a task, the test accuracy of that task will generally drop obviously when the classifier is continually trained on the following tasks. The knowledge acquired from previous tasks are forgotten gradually. It can also be seen that the test accuracy of task 2 and 3 is around $50\%$, which implies the classifier's performance is not better than a random guess. The catastrophic forgetting phenomenon can also be observed in figure \ref{fig:learningcurve}(b) corresponding to the EWC method. Although the ACC has risen from 0.6637 to 0.7307, the classifier still fails to classify the states of task 2 and task 3. Figure \ref{fig:learningcurve}(c) manifests the effectiveness of the GEM method that successfully overcomes catastrophic forgetting. Although 1 epoch of training on task 2 and 3 cannot result in high test accuracy, the continual training of other tasks help to improve the performance of the classifier on task 2 and 3. After the sixth task has been trained, the model's test accuracy on task 2 and 3 is above $86\%$, much better than the other two plots. Every task's test accuracy is higher at the end of training the last task than immediately after training that task, thus the BWT is positive. The numerical results for other 4 task sequences are depicted in figure \ref{fig:allcurves}, which shows similar behaviors.

\section{Conclusion}
\label{sec:conclusion}

Continual learning is a fascinating research topic because it makes artificial intelligence systems learn in a more human-like way. When facing the always changing reality, a machine learning model must be equipped with the capacity of adaptation to new emerging tasks. Previous work \cite{QEWC} has exhibited that a variational quantum classifier can learn a sequence of 3 tasks without catastrophic forgetting. In this work, we move forward to train a variational quantum classifier sequentially on a total of 6 quantum state classification tasks. With resort to the GEM strategy that yields positive knowledge transfer to previous tasks, we have acquired a quantum classifier with better performance. The positive backward transfer explains the improvement brought by GEM. A drawback of GEM is the necessity of computing gradients of previous tasks at each training iteration, while EWC needs to record the diagonal of the Fisher information matrix only at the last training iteration of each task. Another drawback is the episodic memory cost, which would become a heavy burden if the number of tasks is very large. Addressing the two issues will be the focus of our future research. We hope that more exploration and development of quantum continual learning will be done in the future, paving the way to quantum artificial general intelligence.

\end{document}